\documentclass[journal]{IEEEtran}
\usepackage[dvipsnames]{xcolor}
\usepackage{upgreek}
\usepackage{graphicx}
\usepackage{cite}
\usepackage{hyperref}
\usepackage[numbers,sort&compress]{natbib}
\usepackage{multirow}
\usepackage{bm}
\usepackage{orcidlink}
\usepackage{balance}
\ifCLASSINFOpdf
\else
\fi

\title{Ultra-fast Waveguide MUTC Photodiodes over 220 GHz}
\begin{document}

\author{
Linze Li~\orcidlink{0000-0002-6497-3915}, 
Luyu Wang~\orcidlink{0000-0001-9846-6607}, 
Tianyu Long~\orcidlink{0009-0004-9320-6338}, 
Zhouze Zhang~\orcidlink{0009-0003-5002-9696},
Juanjuan Lu~\orcidlink{0000-0003-1938-9119},
BaileChen~\orcidlink{0000-0002-3265-5787},~\IEEEmembership{Senior Member,~IEEE}

\thanks{This work was supported by the National Natural Science Foundation of China under Grant 61975121. The authors are grateful for the device fabrication support from the ShanghaiTech University Quantum Device Lab. (Corresponding author: Baile Chen.)}

\thanks{Linze Li is with the School of Information Science and Technology, ShanghaiTech University, Shanghai 201210, China, with Shanghai Engineering Research Center of Energy Efficient and Custom AI IC, Shanghai 201210, China, with the Shanghai Institute of Microsystem and Information Technology, Chinese Academy of Sciences, Shanghai 200050, China, and also with the University of Chinese Academy of Sciences, Beijing 100049, China (e-mail: lilz@shanghaitech.edu.cn).}

\thanks{Luyu Wang, Tianyu Long, Zhouze Zhang, Juanjuan Lu and Baile Chen are with the School of Information Science and Technology, ShanghaiTech University, Shanghai 201210, China (e-mail: wangly4@shanghaitech.edu.cn; longty2022@shanghaitech.edu.cn; zhangzhz@shanghaitech.edu.cn; lujj2@shanghaitech.edu.cn; chenbl@shanghaitech.edu.cn).}
}

\markboth{}%
{Shell \MakeLowercase{\textit{et al.}}: Bare Demo of IEEEtran.cls for IEEE Journals}
\maketitle

\begin{abstract}
We present InP-based evanescently-coupled waveguide modified uni-traveling carrier photodiodes (MUTC-PDs) exhibiting a breakthrough in bandwidth. The optimization of carrier transport and optical coupling is achieved through a detailed discussion on the design of the cliff layer and waveguide layer. Addressing the parasitic capacitance challenge, we introduce benzocyclobutene (BCB) beneath the PD electrodes, effectively overcoming the bandwidth bottleneck associated with the RC time constant.
Devices with sizes of 2 $\times$ 7 $\upmu$m$^{2}$ and 2$\times$ 10 $\upmu$m$^{2}$ achieve
3-dB bandwidths over 220 GHz, along with external responsivities of 0.161 A/W and 0.237 A/W, respectively. Notably, the RF output power reaches a peak of -1.69 dBm at 215 GHz for 2 $\times$ 15 $\upmu$m$^{2}$ PDs. 

\end{abstract}

\begin{IEEEkeywords}
Photodiodes (PD), high-speed photodetectors, modified uni-traveling-carrier (MUTC), THz generation.
\end{IEEEkeywords}

\IEEEpeerreviewmaketitle

\section{Introduction}

\IEEEPARstart{T}{he} pursuit of ultra-high-speed data transmission has been one of the primary driving forces for research on ultra-fast photodetectors. The advancement of communication frequencies into the terahertz (THz) domain (0.1-1 THz) brings forth significant challenges, particularly in the generation and detection of ultra-high-frequency signals for analog and digital photonic applications \cite{analog-digital}. In these systems, ultra-fast photodiodes (PDs) play a vital role as the key component, typically determining the maximum allowable operating frequency and dynamic range \cite{jinwei300}. For the generation of THz waves, photonic techniques that based on high-speed photodetectors are considered to be superior to conventional electronic techniques with respect to wide frequency bandwidth, tunability, and stability \cite{nagatsuma2009high}. Meanwhile, for the receiver end of the digital communication systems, integrating ultra-broadband photodetectors is essential for the realization of next-generation 200 GBaud interconnects \cite{beckerwerth2023high,umezawa2022ultra}. Among various types of high-speed photodiodes, uni-traveling carrier photodiodes (UTC-PDs) \cite{UTC2000,UTC2020}, utilizing only electrons as the active carriers in the drift layer, stand out as highly promising structures for applications demanding both high speed and high power \cite{UTC_highpower}.

 Surface-illuminated PDs are capable of achieving impressive speed and saturation characteristics. The back illuminated modified UTC-PDs (MUTC-PDs) reported in \cite{qinghua230} exhibited a wide bandwidth of 230 GHz by inductance peaking and reducing the diameter of PDs down to 3 $\upmu$m. MUTC-PDs flip-chip bonded onto the diamond submount realized RF output power of -3 dBm at 150 GHz and -5.7 dBm at 165 GHz \cite{xiaojun150}. However, due to the vertical nature of carrier transport and optical coupling, surface-illuminated PDs face challenge in simultaneously achieving short carrier transit time and high responsivity \cite{coupling}. 
In contrast, waveguide PDs (WG-PDs) offer a promising solution to the bandwidth-responsivity tradeoff by coupling light laterally into the active region of the device through an optical waveguide. 
E. Rouvalis et al. reported evanescent-coupled UTC-PDs with a 170 GHz bandwidth and a responsivity up to 0.27 A/W \cite{b4}. A waveguide integrated PD module with a 3dB-bandwidth of 145GHz and a responsivity of 0.4 A/W has been demonstrated \cite{b8}.

In this work, we demonstrated evanescently-coupled waveguide MUTC-PDs with ultra-fast speed and high output power. 
The cliff layer doping level was carefully designed to enhance the electron transport and compensate for space charge effect under high optical power injection.
Evanescently optical coupling was also optimized by tuning the refractive index and thickness of the waveguide layer. Furthermore, for better RC-limited bandwidth, we investigated the primary factors contributing to the parasitic capacitance of the devices. By implementing benzocyclobutene (BCB) under the coplanar waveguide (CPW) electrodes, such capacitance was effectively reduced compared to the previous work \cite{130GHz}. 
To the best of our knowledge, this work obtained a record-high 3-dB bandwidth exceeding 220 GHz for 2 $\times$ 7 $\upmu$m$^{2}$ and 2 $\times$ 10 $\upmu$m$^{2}$ devices and a high saturation output power of -1.69 dBm at 215 GHz for 2 $\times$ 15 $\upmu$m$^{2}$ device.


\section{Design and Fabrication of Device}

\begin{figure}
    \centering
    \includegraphics[width=\linewidth]{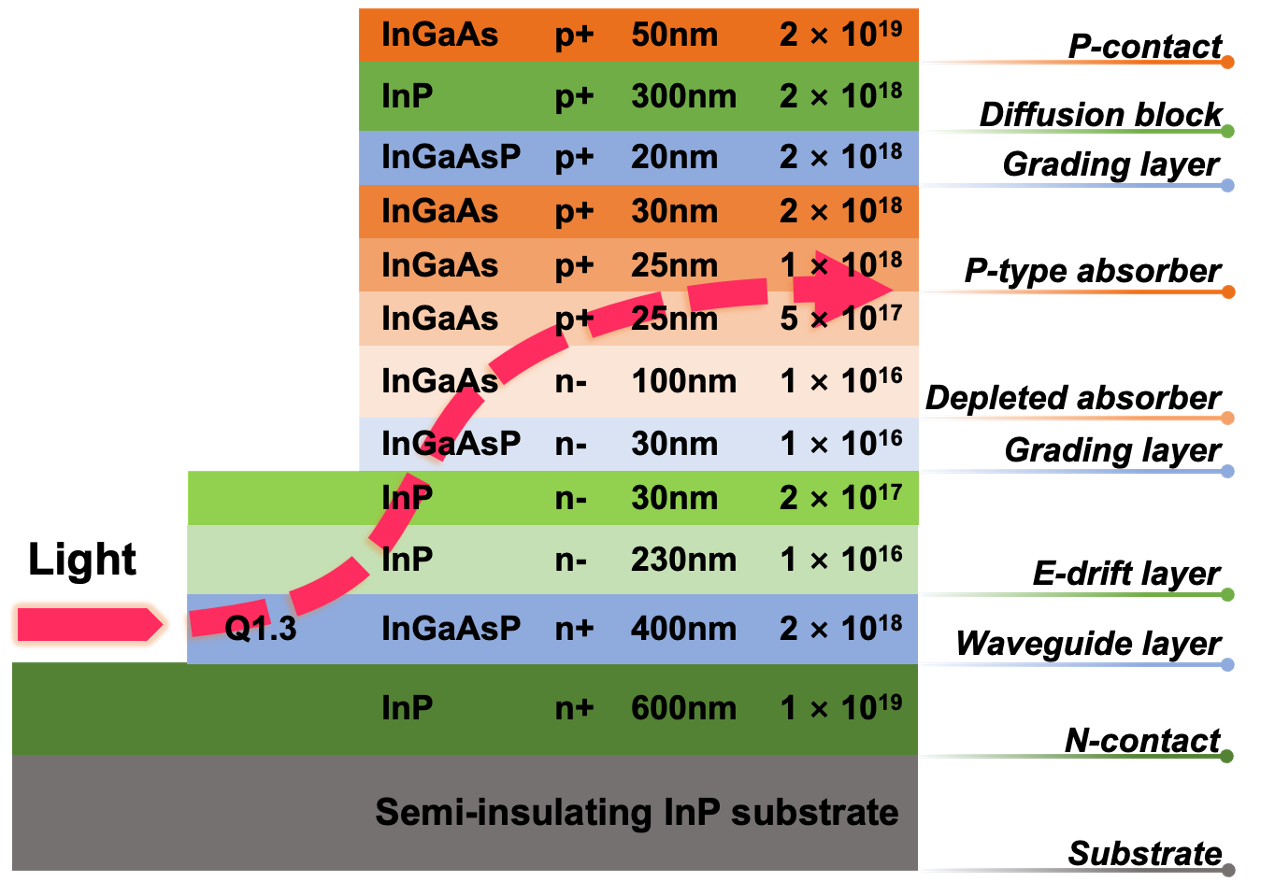}
    \caption{Epi-layer structure of the InGaAs/InP photodiode.}
    \label{fig1}
\end{figure}

\subsection{MUTC Structure Design}

The designed epitaxial structure of the InGaAs/InP wafer was grown on the semi-insulating InP substrate, as shown in Fig. \ref{fig1}. The 180 nm absorption layer was partially depleted to build a high electric field across the hererojunction interface. An 80 nm p-doped InGaAs undepleted absorption layer with a step-grading profile(5 × 10$^{17}$ cm$^{-3}$,1 × 10$^{18}$ cm$^{-3}$, and 2 × 10$^{18}$ cm$^{-3}$) was implemented to form a quasi-electric field which assists electron transport. Additionally, a 250 nm slightly n-doped InP drift layer was designed to serve as the charge compensation layer. Heavily p-doped InGaAs and heavily n-doped InP layers were deposited as the p and n contact,respectively. InGaAsP quaternary layers were grown at InGaAs/InP heterojunction interfaces to smooth the energy band discontinuities. 

\begin{figure}
    \centering
    \includegraphics[width=\linewidth]{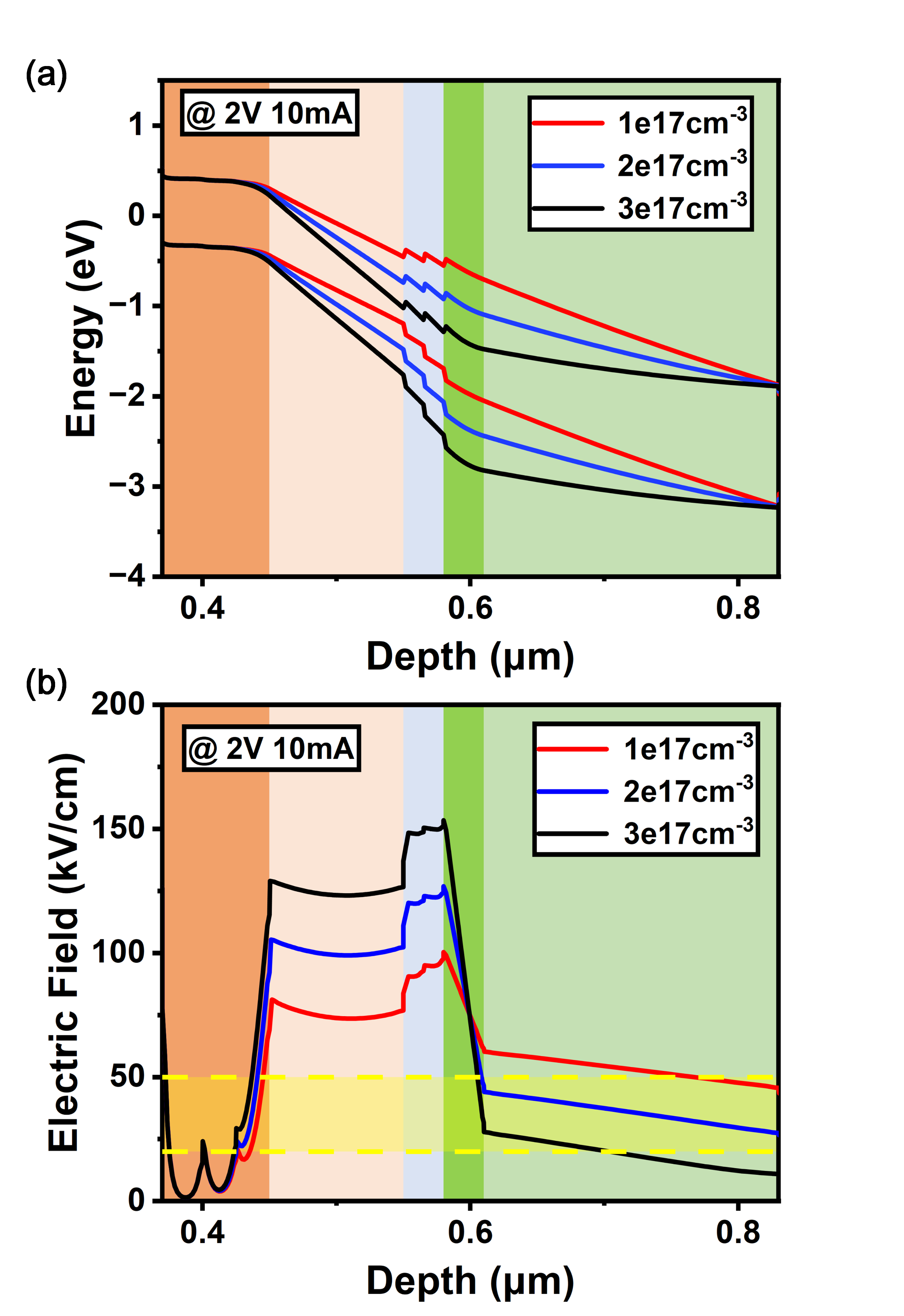}
    \caption{Simulated (a) energy band diagram and (b) electric field distribution for the 2 $\times$ 15 $\upmu$m$^{2}$ device with different cliff layer doping levels at -2V bias and 10 mA photocurrent.}
    \label{fig2}
\end{figure}

A 30 nm cliff layer with the optimized doping level of 2 × 10$^{17}$ cm$^{-3}$ was inserted in the drift layer. In order to investigate the influence of the cliff layer with different doping level, we simulated the band diagram and electric field distribution as the doping concentration of the cliff layer varies from 1 × 10$^{17}$ cm$^{-3}$ to 3 × 10$^{17}$ cm$^{-3}$. Simulations are performed by the Silvaco TCAD software at -2V bias and 10 mA photocurrent for the 2 $\times$ 15 $\upmu$m$^{2}$ device. Increasing the doping concentration in the cliff layer can significantly decrease the conduction band barrier across the heterogeneous interface in the depletion region, as shown in Fig.\ref{fig2} (a). This favors electrons transport and alleviate the accumulation of electrons at the InGaAs/InP interfaces. In addition, as depicted in Fig. \ref{fig2} (b), the higher electric field induced by higher doping level of cliff layer can also help electrons to cross the heterogeneous inter-facial barriers and mitigate the effect of space charge effect at a high current level \cite{lizhi}. However, when the n-doping of the cliff layer increases to 3 × 10$^{17}$ cm$^{-3}$, the electric field will be too low to sustain the fast transport of electrons in InP drift layer. A doping level of 2 × 10$^{17}$ cm$^{-3}$ can maintain an electric field of 20-50 kV/cm in the whole e-drift layer to enable electron overshoot \cite{Qinghua2} and ensure a short carrier transit time. 

\begin{figure}
    \centering
    \includegraphics[width=\linewidth]{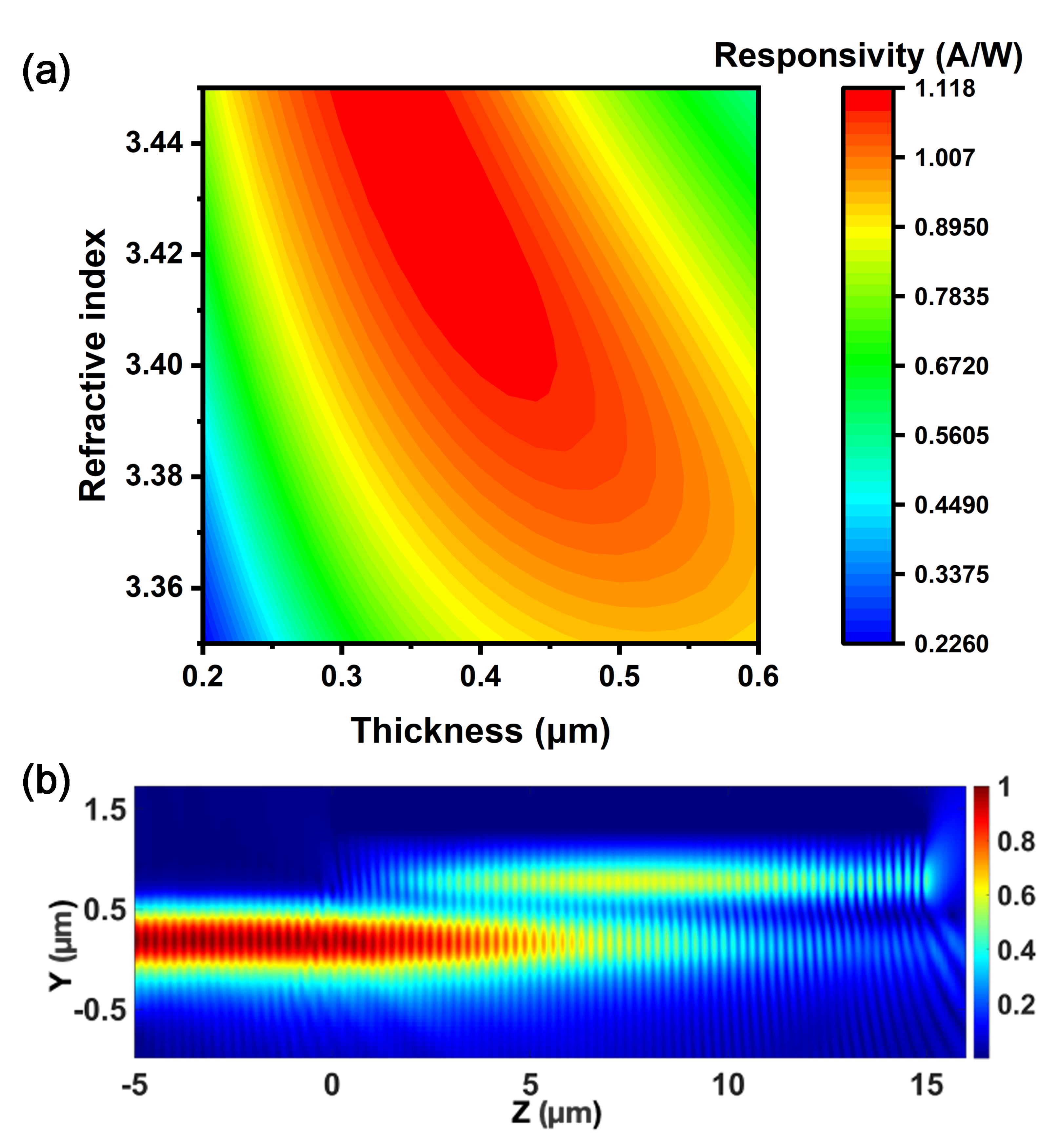}
    \caption{(a) Simulated internal responsivity versus refractive index and thickness of the waveguide layer for the 2 $\times$ 15 $\upmu$m$^{2}$ device. (b) Simulated optical intensity as light propagates through the structure. }
    \label{fig3}
\end{figure}

\subsection{Optical coupling design}

A single-layer InGaAsP waveguide is designed for optical coupling and the InP drift layer above acts as the cladding for it. For better saturation characteristics, longer coupling length is preferred to allow more uniform evanescently coupling of light from the passive waveguide to the active region of the photodiode. However, longer PD length would result in a higher junction capacitance, degrading the RC-limited bandwidth. To balance such trade-off, the largest length of the device is set to be 15 um. In this work, the internal responsivity of the 2 $\times$ 15 $\upmu$m$^{2}$ device was simulated with respect to different refractive index and thickness of the waveguide layer, as shown in Fig. \ref{fig3} (a). An optimal responsivity of 1.1 A/W was achieved when the thickness and refractive index of the waveguide layer is 0.4 $\upmu$m and 3.41 (Q 1.3 InGaAsP) respectively. Fig. \ref{fig3} (b) shows the corresponding optical intensity as light propagates through the device.

\subsection{Fabricaiton}

\begin{figure}
    \centering
    \includegraphics[width=0.75\linewidth]{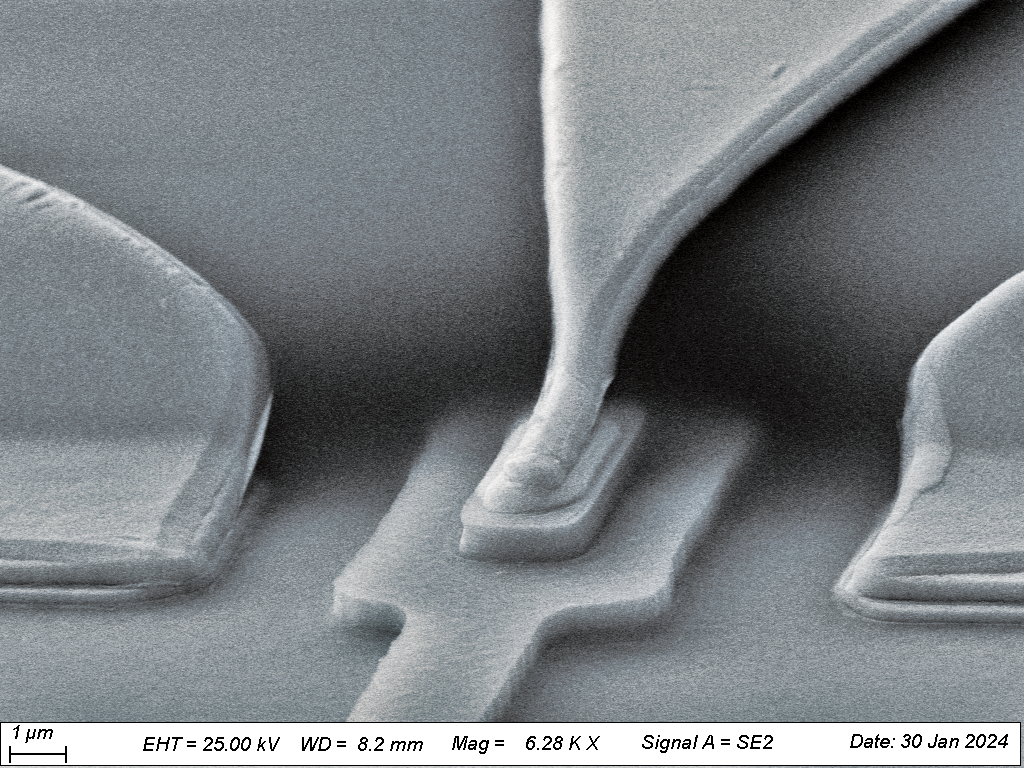}
    \caption{Front view SEM image of the fabricated waveguide MUTC-PD.}
    \label{fig4}
\end{figure}

The device was fabricated into a triple-mesa structure, as shown in Fig. \ref{fig4}. Inductively coupled plasma (ICP) dry etch process was conducted to achieve vertical sidewalls and precise control of the size of p-mesa and waveguide mesa, while wet etch (H$_{3}$PO$_{4}$:HCl = 1:3) was used for electrical isolation between n-mesas. Deposition of Ti/Pt/Au/Ti and GeAu/Ni/Au as p-contact and n-contact was followed by the rapid thermal annealing process at 360 $^{\circ}$C to achieve the low contact resistance. 

\begin{figure}
    \centering
    \includegraphics[width=\linewidth]{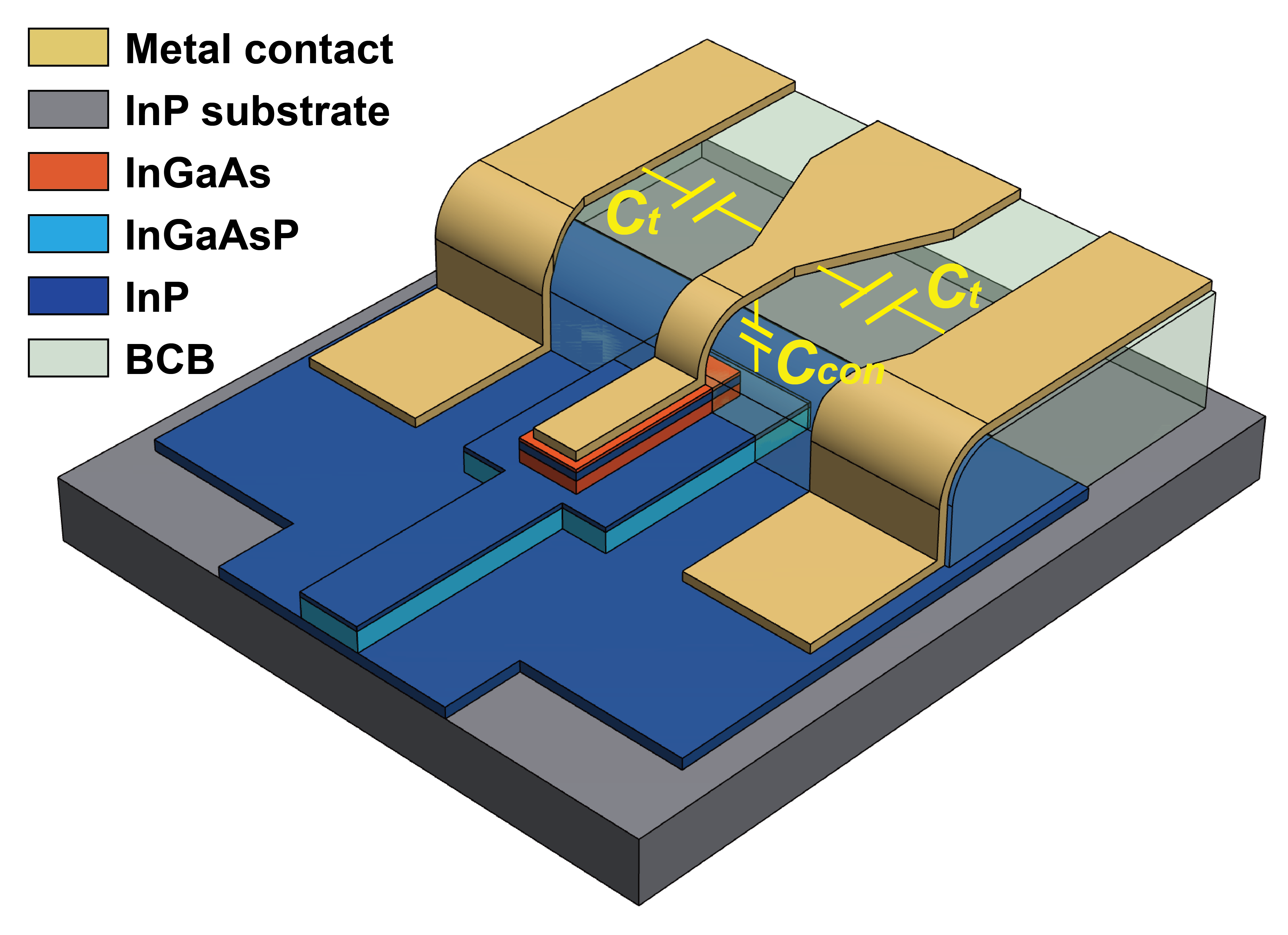}
    \caption{Schematic diagram dipicting the main components of the parasitic capacitance in the waveguide MUTC-PD. }
    \label{fig5}
\end{figure}

In our previous work \cite{130GHz}, the parasitic capacitance is the major limiting factor of RC response, which mainly consists of C$_{con}$ and C$_{t}$, as shown in Fig. \ref{fig5}. C$_{con}$ is the capacitance between the CPW signal pad and the n-contact layer \cite{Ccon}. Considering the stability of the air-bridge structure, the spacing distance between these metals is limited to be around 1 $\upmu$m, leading to a large C$_{con}$. For a better RC limited bandwidth, 4 $\upmu$m thick Benzocyclobutene (BCB) was introduced under the CPWs (see Fig. 3) to eliminate the need for the previous air-bridge structure and effectively pad the signal electrode \cite{German}. Such a non-suspended structure can also ensure stable and uniform contact between p-mesa and CPWs for devices with width down to 2 $\upmu$m.

\begin{figure}
    \centering
    \includegraphics[width=\linewidth]{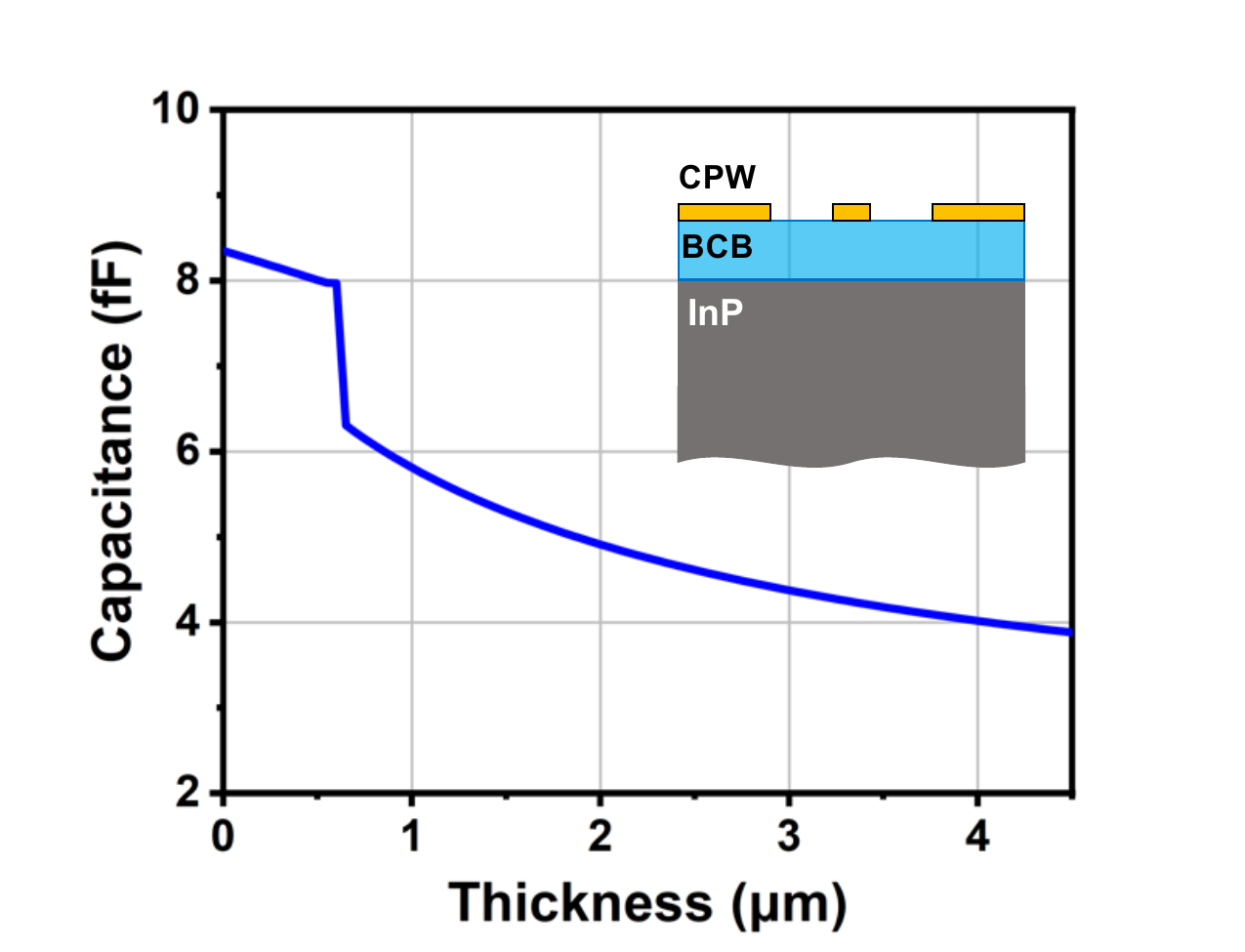}
    \caption{Simulated curve of C$_{t}$ versus thickness of BCB on InP substrate. Insert is the schematic diagram of simulated model.}
    \label{fig6}
\end{figure}

C$_{t}$ is the capacitance introduced between electrodes of the CPW itself. The permittivity and thickness of the multilayered substrates have a significant impact on C$_{t}$ \cite{BCBtheory,BCBtheory2}. As shown in Fig. \ref{fig6}, We employed Computer Simulation Technology (CST) EM Studio to simulate the curve of C$_{t}$ versus thickness of BCB on InP substrate. Due to the 4 $\upmu$m thick BCB strip-supporting layer, the C$_{t}$ is greatly reduced by more than half. 

\section{Characterization and Discussion}

\subsection{DC and Optical Characteristics}
\begin{figure}
    \centering
    \includegraphics[width=1\linewidth]{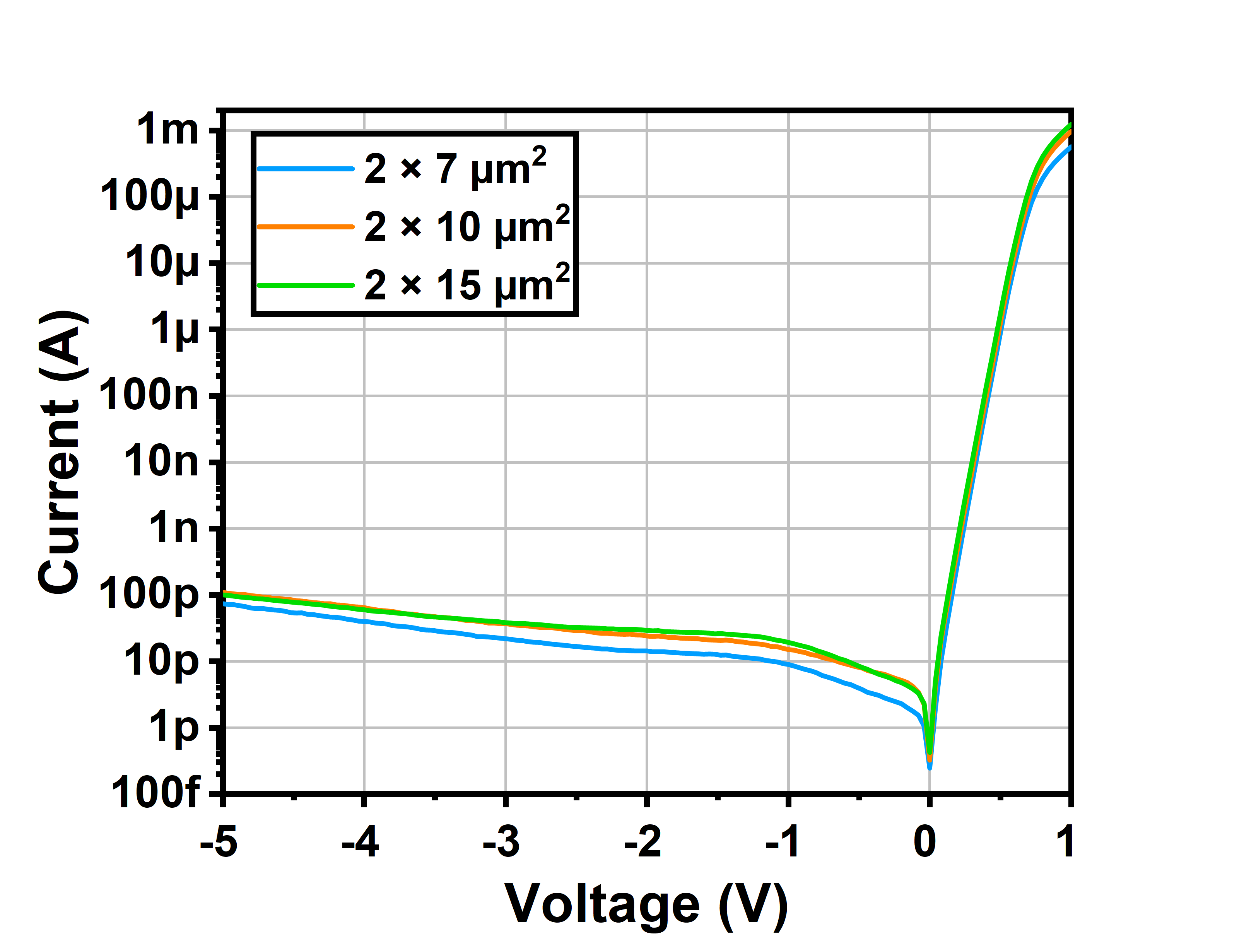}
    \caption{Dark current versus bias voltage characteristic for the devices with different lengths.  }
    \label{fig7}
\end{figure}

The dark current versus bias voltage characteristics for devices with different lengths are shown in Fig. \ref{fig7}. A typical dark current is around 100 pA under -5 V bias, which is at least one order of magnitude lower 
 than other InGaAs/InP based MUTC-PDs \cite{qinghua230,xiaojun150,German}. Such low dark current can be attributed to the careful protection of the PD sidewalls and waveguide mesa during the dry etch process.

\begin{figure}
    \centering
    \includegraphics[width=1\linewidth]{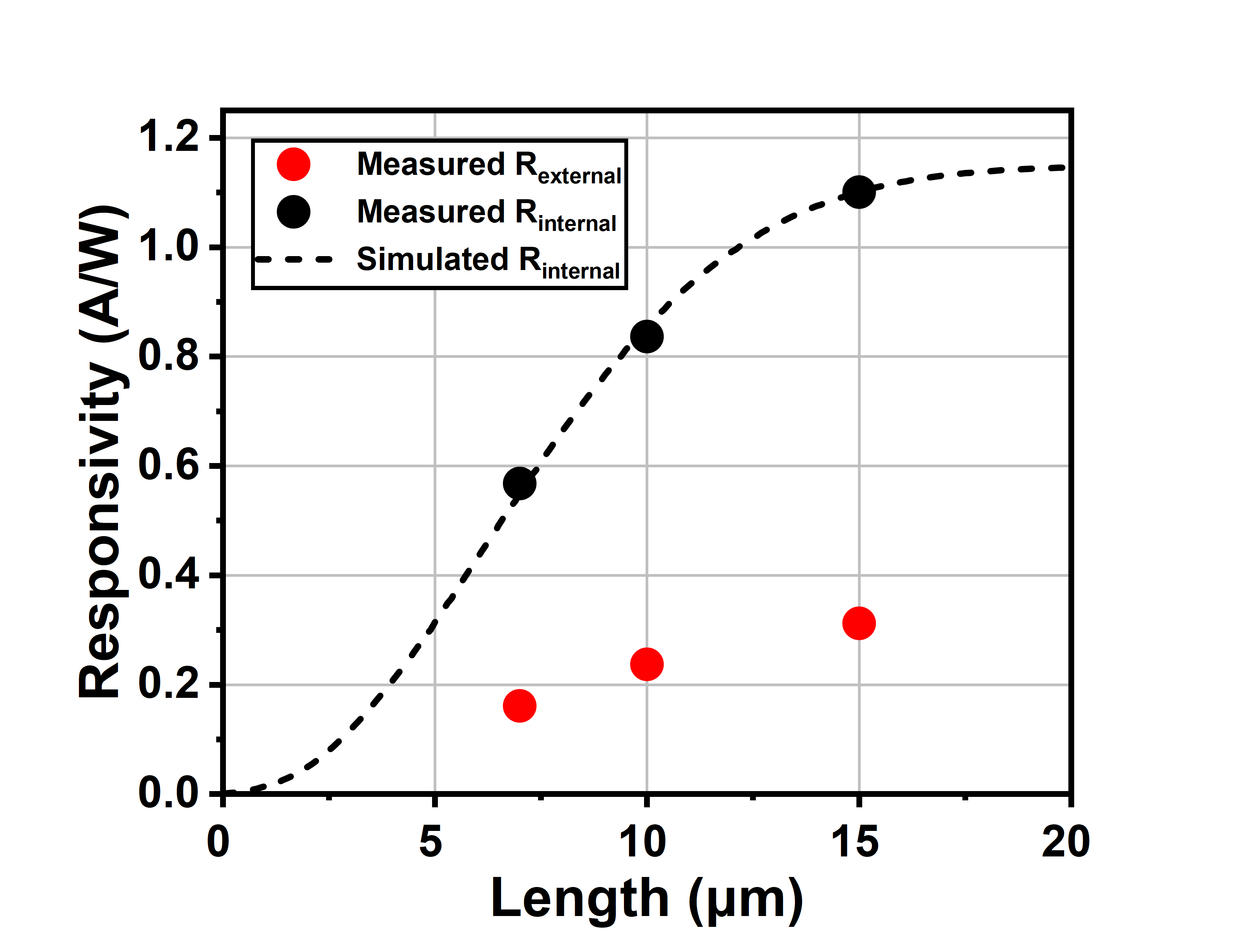}
    \caption{Measured external (red circles) and internal (black circles) responsivities and simulated internal responsivities (dashed line) of the devices with different lengths.}
    \label{fig8}
\end{figure}

To calculate the internal responsivity of the devices, the fiber-to-waveguide coupling loss and waveguide propagation loss were experimentally measured to be 5.5 dB in total.
The assessment of fiber-to-waveguide coupling loss involved utilizing lensed fiber to couple light into and out of the test waveguide, followed by a comparison of input and output optical power. The test waveguide featured a slight curve near the waveguide facet to avoid collecting optical power outside the waveguide. 
Propagation loss was evaluated by studying the responsivity of devices with varying waveguide lengths.
In Fig \ref{fig8}, the external and internal responsivities of devices with different lengths are illustrated. Devices with lengths of 7 $\upmu$m, 10 $\upmu$m and 15 $\upmu$m show external responsivities of 0.161 A/W, 0.237 A/W and 0.312 A/W at 1550 nm wavelength. 
The associated internal responsivities are found to be 0.567 A/W, 0.835 A/W, and 1.099 A/W, respectively. The simulated internal responsivity versus the length of PDs using Ansys Lumerical FDTD is represented by the dashed line, demonstrating a robust agreement with the measured results.

\subsection{RF Characteristics}

\begin{figure}
    \centering
    \includegraphics[width=1\linewidth]{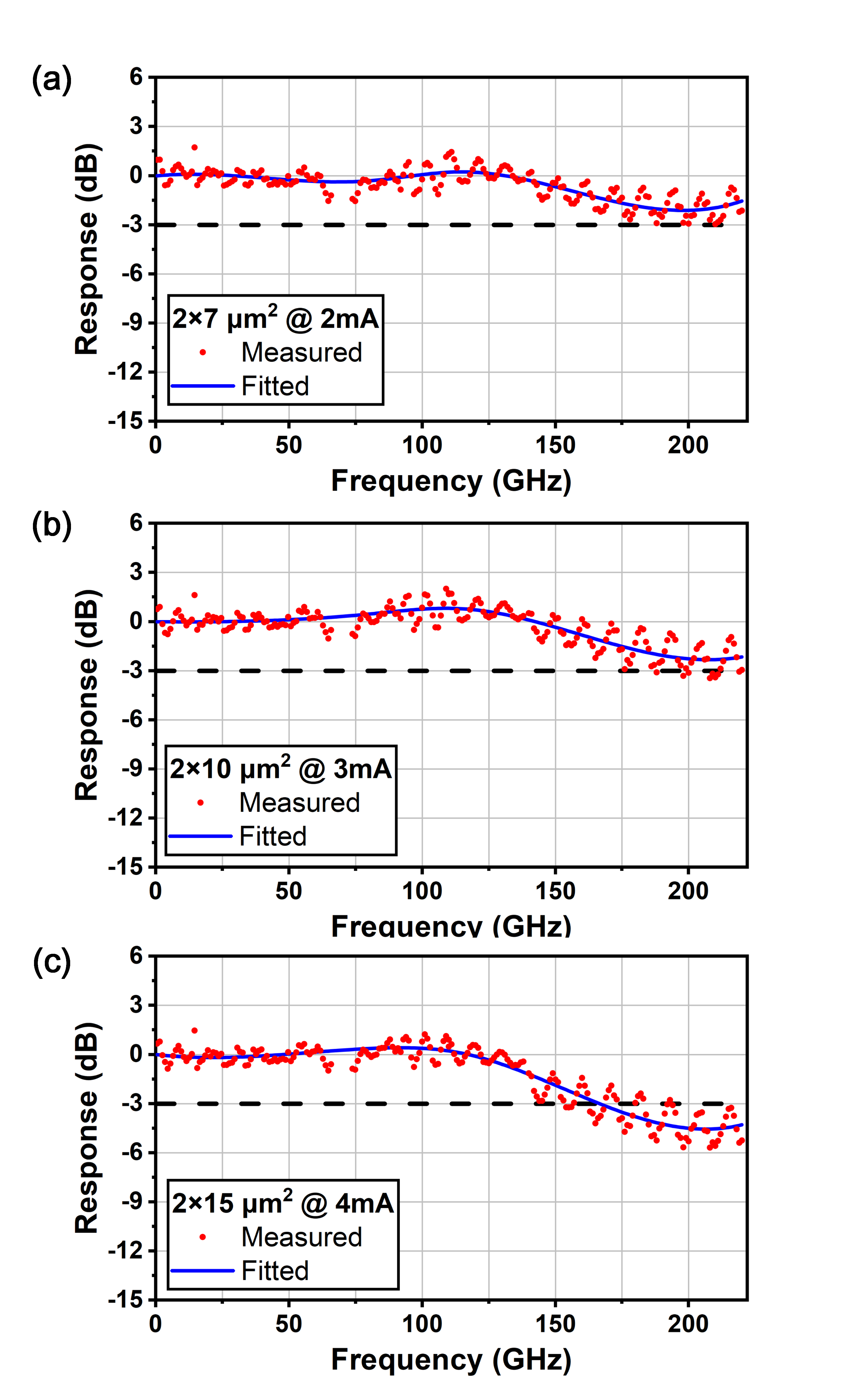}
    \caption{Frequency response of the (a) 2 $\times$ 7 $\upmu$m$^{2}$ device at 2mA photocurrent, (b) 2 $\times$ 10 $\upmu$m$^{2}$ device at 3mA photocurrent and (c) 2 $\times$ 15 $\upmu$m$^{2}$ device at 4mA photocurrent under -1.2 V bias.}
    \label{fig9}
\end{figure}

The frequency response and the saturation performance of the InGaAs/InP waveguide devices were investigated by an optical heterodyne setup \cite{2um}. Two tunable lasers with wavelengths near 1.55 $\upmu$m were combined to generate a heterodyne beat note with a wide frequency tuning range from Megahertz to Terahertz. A lensed fiber with a 2-$\upmu$m spot diameter was used to couple the modulated optical signal into the cleaved waveguide facet with no anti-reflective coating. The output RF power was obtained by two measurement setups. From DC to 110 GHz, the RF signal was collected by the Rohde \& Schwarz power meter (NRP-Z58) through the GSG Model 110H probe. Frequency-dependent loss value introduced by cables, bias tee and GSG probe was calibrated by a vector network analyzer (VNA) with corresponding calibration kit (85058E \& CS-5 for DC to 67 GHz, WR10-VDI \& CS-5 for 73.8 GHz to 110 GHz). For frequency above 110 GHz, a VDI power meter (PM5) and a set of waveguide probes covering 110 GHz to 220 GHz were used to measure the THz power generation. The loss of the probes and waveguide connectors provided by the manufacturer was de-embedded from the measurement results.

In Fig. \ref{fig9}, the frequency response of devices with lengths of 7$\upmu$m, 10$\upmu$m, and 15$\upmu$m is depicted at a -1.2 V bias voltage, corresponding to photocurrents of 2 mA, 3 mA, and 4 mA, respectively. The 2 $\times$ 7 $\upmu$m$^{2}$ and 2 $\times$ 10 $\upmu$m$^{2}$ devices showcase ultra-fast speeds, with a power drop of less than 3 dB from DC to 220 GHz, indicating a 3-dB bandwidth exceeding 220 GHz. Measurement beyond 220 GHz was not conducted due to limitations in the current lab setup. The 2 $\times$ 15 $\upmu$m$^{2}$ device exhibits a lower bandwidth of 165 GHz, constrained by the RC response owing to a larger junction capacitance.

\begin{figure}
    \centering
    \includegraphics[width=1\linewidth]{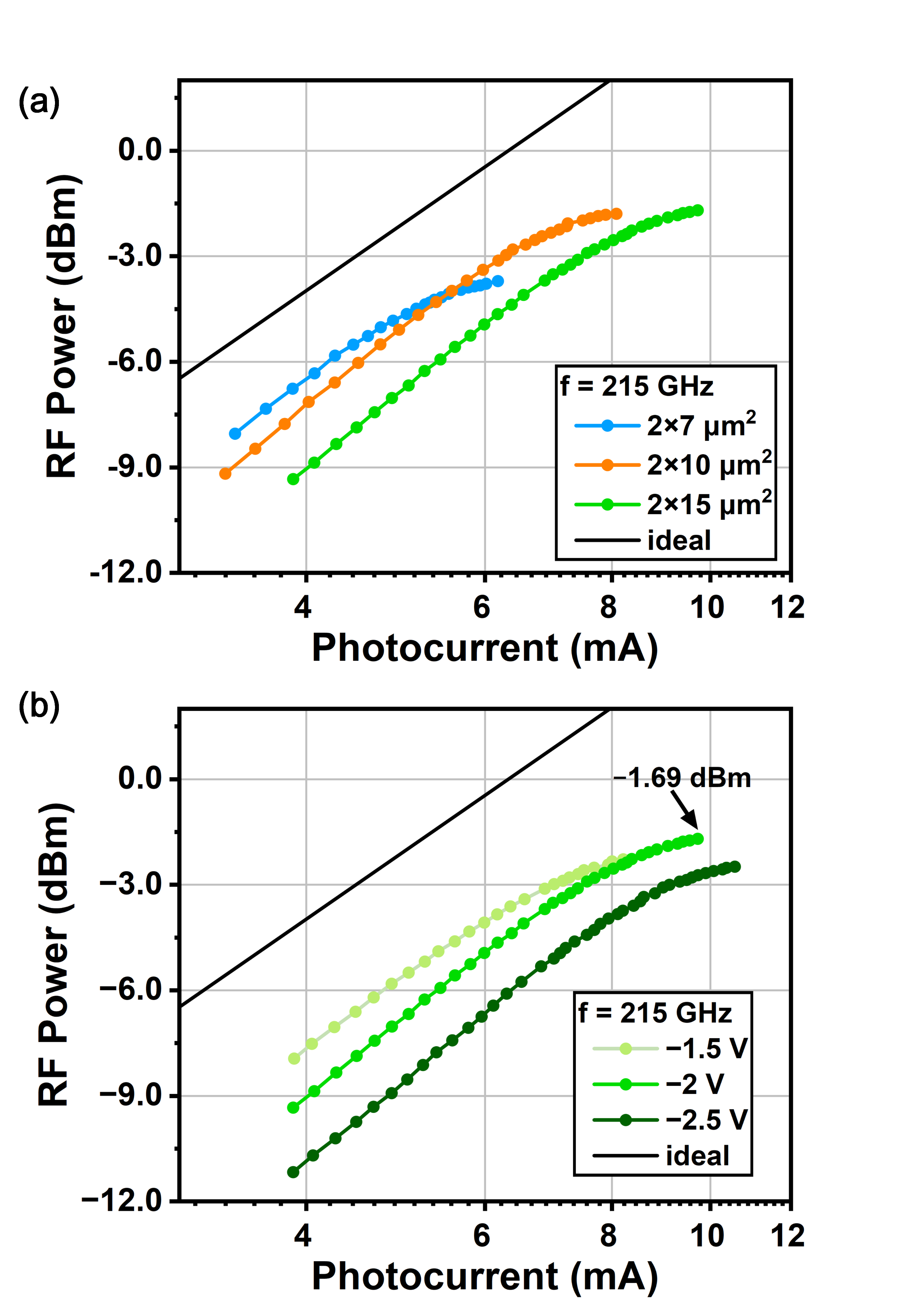}
    \caption{(a) RF power versus photocurrent for the waveguide MUTC-PDs with different lengths under -2V bias. (b) RF power versus photocurrent for the 2 × 15 $\upmu$m$^{2}$ device under different bias voltages. The measurement frequency is fixed at 215 GHz.}
    \label{fig10}
\end{figure}

At a fixed frequency of 215 Ghz, the saturation performance of PDs with various lengths is characterized by measuring the RF output power as a function of the average photocurrent, as shown in Fig. \ref{fig10} (a). Under -2 V bias, devices with lengths of 7$\upmu$m, 10$\upmu$m and 15$\upmu$m reach output power of -3.70 dBm, -1.79 dBm and -1.69 dBm at 215 GHz, respectively. For photocurrent below 5 mA, shorter devices output higher RF power due to better 3dB bandwidth. However, the 2 $\times$ 7 $\upmu$m$^{2}$ device saturates earlier at a photocurrent of 6.2 mA, leading to overall lower RF output power. Such a phenomenon is due to higher current densities as well as the in-homogeneous light intensity distribution in small-sized devices. The RF output power versus the average photocurrent of the 2 $\times$ 15 $\upmu$m$^{2}$ device under different bias voltages is shown in Fig. \ref{fig10} (b). At low light intensities, the difference between the RF power and the ideal power decreases with increasing light intensity, which can be attributed to faster electron transport resulting from self-induced quasi-field in the absorber \cite{selfinduce}. At higher light intensities, saturation occurs due to growing charge accumulation. The saturation current increases from 8.2 to 10.6 mA when the bias voltage increases from 1.5 to 2.5 V, since higher electric field can suppress the space charge effect. The device exhibits the best saturation performance under -2 V bias, with a 1-dB compression point at 10.6 mA and output power of -1.69 dBm. Under -2.5 V, the electric field in the drift layer is too high to enable electron’s overshoot and result in a lower bandwidth and output power. 

\subsection{S11 Parameter Fitting}

\begin{figure}
    \centering
    \includegraphics[width=1\linewidth]{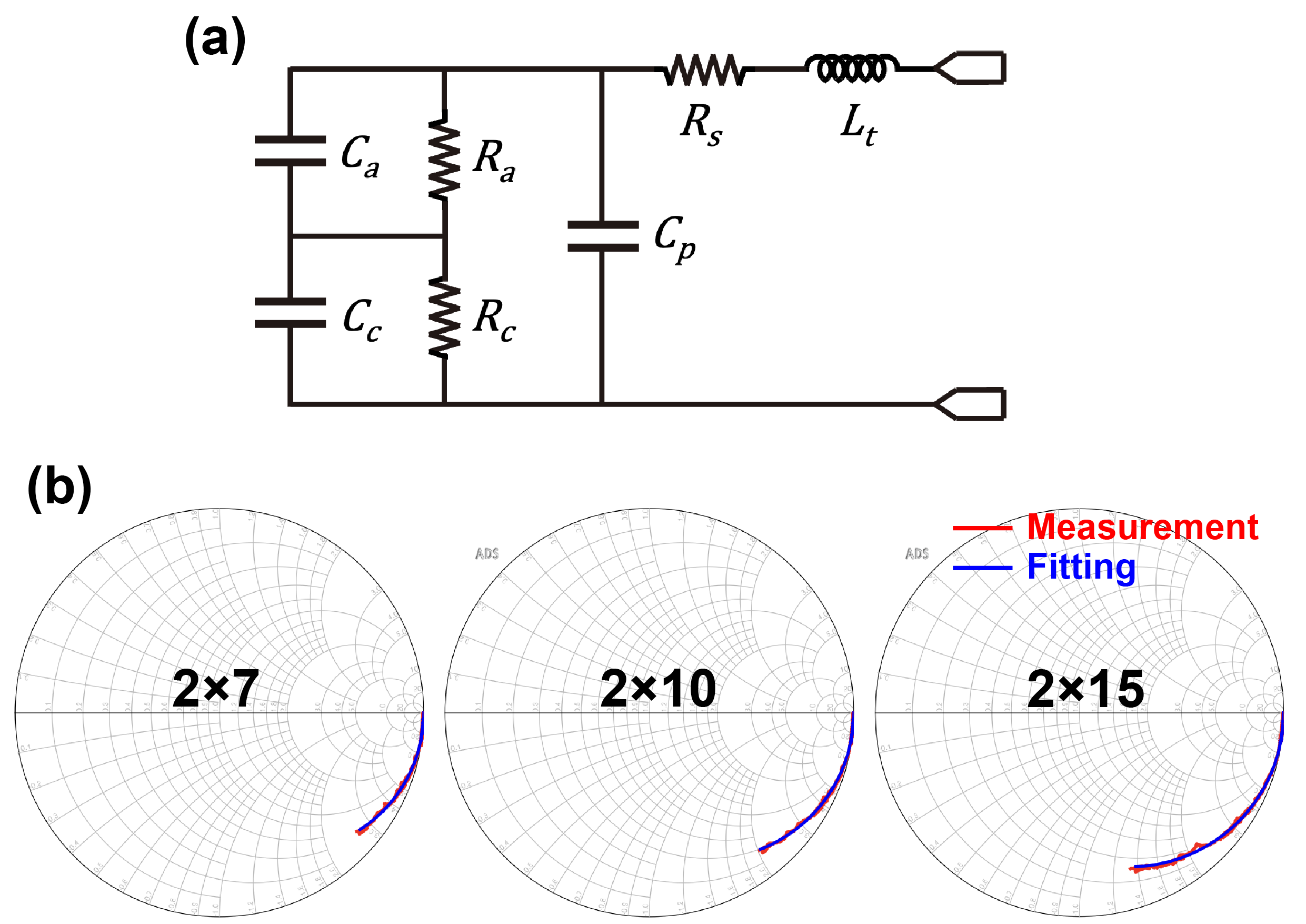}
    \caption{(a) Equivalent circuit model of the MUTC devices for S11 fitting. (b) Measured (blue line) and fitted (red line) S11 data of 2 × 7 $\upmu$m$^{2}$, 2 × 10 $\upmu$m$^{2}$ and 2 × 15 $\upmu$m$^{2}$ devices with 50 MHz-110 GHz frequency range under -1.2 V bias.  }
    \label{fig11}
\end{figure}

The S11 parameters up to 110 GHz were acquired through on-wafer network analyzer measurements. These parameters allows for the comprehensive characterization of the impedance profile of the photodetectors. To extract the intrinsic equivalent circuit parameters, we employed parameter fitting within the Advanced Design System (ADS) software. The equivalent circuit model illustrated in Fig. \ref{fig11} (a) was used to represent the waveguide PDs. The model incorporates elements representing the series resistance \( R_{s} \), CPW inductance \( L_{t} \), and the parasitic capacitance \( C_{p} \). \( C_a \) and \( C_c \) were designated as the junction capacitance of the depleted InGaAs absorber and the InP collector, while \( R_a \) and \( R_c \) represented the junction resistances, correspondingly.

The alignment of the measured S11 (red line) with the fitted results (blue line) on the Smith Charts, as shown in Fig. \ref{fig11} (b), validated the accuracy of the circuit model. The extracted parameters are given in Table I. Notably, The Smith chart shows that the S11 values start at the open circuit point, indicating a high junction resistance (\( R_a \) and \( R_c \)), which is set to be 100 M\(\Omega\) and not presented in the table. \( C_{c} \) and \( C_{a} \) were calculated separately using a parallel plate capacitor approximation, reflecting the dual-material depletion zones characteristic of the MUTC-PD. 

The total parasitic capacitance \( C_{p} \)	, which includes \( C_{t} \), \( C_{con} \), and other parasitic elements, maintained a consistent value of about 4.5 fF across the devices. In contrast to our earlier work where parasitic capacitance were approximately 20 fF \cite{130GHz}, the current study demonstrates a pronounced decline to 4.5 fF. This significant reduction is largely owing to the application of BCB under the electrodes, contributing substantially to the enhanced bandwidth.

\begin{table}[t]
\caption{Fitting parameters}
\label{table:your_table_label}
\centering
\begin{tabular}{|c|c|c|c|c|c|}
\hline
Area ($\upmu$m$^{2}$) & Ca (fF) & Cc (fF) & Rs ($\Omega$) & Cp (fF) & Lt (pH) \\ \hline
2 × 7  & 17.2 & 5.5 & 23.3 & 4.4  & 49.9 \\ \hline
2 × 10 & 24.6 & 7.8 & 19.6 & 4.56 & 51.7 \\ \hline
2 × 15 & 36.9 & 11.8 & 16.1 & 4.38 & 57.7 \\ \hline
\end{tabular}
\end{table}

\begin{figure}
    \centering
    \includegraphics[width=1\linewidth]{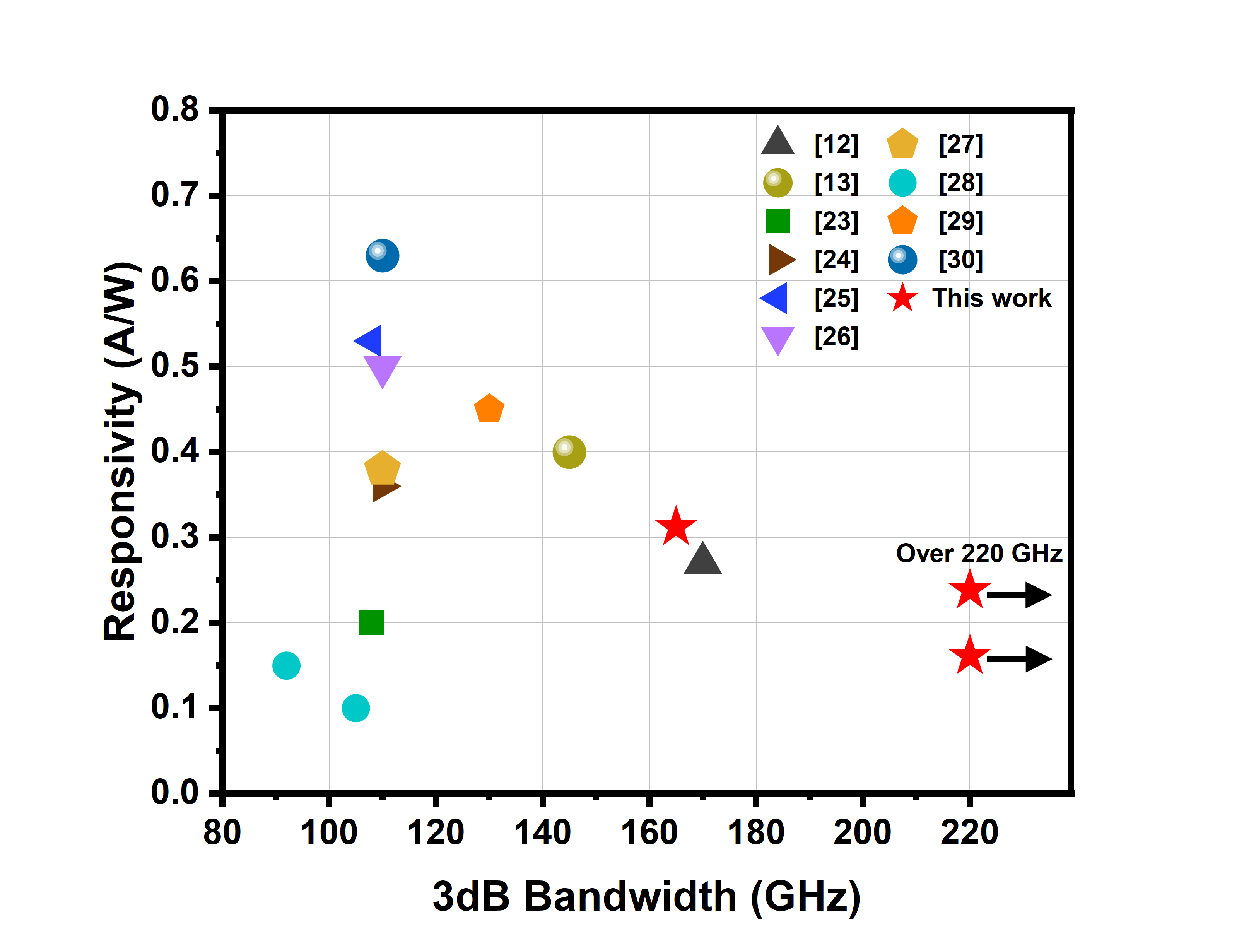}
    \caption{3dB bandwidth and responsivity of waveguide coupled PDs based on group III-V semiconductors reported in literature \cite{b1,b2,b3,b4,b5,b6,b7,b8,b9,b10}.}
    \label{fig12}
\end{figure}

Fig. \ref{fig12} summarize the state-of-the-art responsivity versus 3-dB bandwidth of InGaAs/InP based WG-PDs. Our work shows the highest bandwidth along with a good responsivity. Given that only a single waveguide layer is utilized for optical coupling in this study, the incorporation of a diluted waveguide and a spot-size converter could further enhance coupling efficiency.

\section{Conclusion}
In this work, we have reported InGaAs/InP based ultra-fast waveguide MUTC-PDs operating at 1550 nm wavelength. The cliff layer doping level, as well as the refractive index and thickness of the waveguide layer, are meticulously adjusted to guarantee favorable RF and optical characteristics. Meanwhile, BCB is introduced to reduce the parasitic capacitance and futher extend the bandwith of PDs. The 3 dB bandwidths of the 2 $\times$ 7 $\upmu$m$^{2}$ and 2 $\times$ 10 $\upmu$m$^{2}$ devices surpass 220 GHz under -1.2 V bias. At 215 GHz, an output RF power of -1.69 dBm is measured, corresponding to a saturation current of 10.6 mA. These promising findings suggest the potential for THz generation through InP photonics integrated circuits.






\small
\bibliographystyle{IEEEtran}
\bibliography{ref}






\end{document}